\documentclass[twocolumn,groupedaddress,
superscriptaddress,preprintnumbers,
amsmath,amssymb,
aps,
]{revtex4}
\usepackage{graphicx}
\usepackage{dcolumn}
\usepackage{bm}%
\usepackage[colorlinks=true,citecolor=blue]{hyperref}
\hypersetup{colorlinks=true,citecolor=blue,linkcolor=red,urlcolor=blue}

\usepackage{float}
\usepackage{amsmath}

\usepackage{color}
\definecolor{Green}{RGB}{0,204,102}
\definecolor{Purple}{RGB}{102,0,255}
\definecolor{Blue}{RGB}{51,153,255}
\definecolor{Red}{RGB}{151,010,010}

\makeatletter
\def\@bibdataout@aps{%
\immediate\write\@bibdataout{%
@CONTROL{%
apsrev41Control%
\longbibliography@sw{%
    ,author="08",editor="1",pages="1",title="0",year="1"%
    }{%
    ,author="08",editor="1",pages="1",title="",year="1"%
    }%
  }%
}%
\if@filesw \immediate \write \@auxout {\string \citation {apsrev41Control}}\fi 
}
\makeatother

\begin{document}

\title{ Natural hyperbolicity in the layered hexagonal crystal structure}
\author{Ali Ebrahimian}
\email{aliebrahimian@ipm.ir}
\affiliation{School of Nano Science, Institute for Research in Fundamental Sciences (IPM), Tehran 19395-5531, Iran}
\author{Reza Asgari}
\email{asgari@ipm.ir}
\affiliation{School of Physics, Institute for Research in Fundamental Sciences (IPM), Tehran 19395-5531, Iran}
\affiliation{ARC Centre of Excellence in Future Low-Energy Electronics Technologies, UNSW Node, Sydney 2052, Australia}
\date{\today}

\begin{abstract}
Discovering the physical requirements for meeting the indefinite permittivity in natural material as well as proposing a new natural hyperbolic media offer a possible route to significantly improve our knowledge and ability to confine and controlling light in optoelectronic devices. We demonstrate the hyperbolicity in a class of materials with hexagonal P6/mmm and P6$_3$/mmc layered crystal structures and its physical origin is thoroughly investigated. By utilizing density functional theory and solving the Bethe-Salpeter equation (BSE), we find that the layered crystal structure and symmetry imposed constraints in Li$_3$N gives rise to an exceedingly strong anisotropy in optical responses along in- and out-of-plane directions of the crystals making it a natural hyperbolic in a broad spectral range from the visible spectrum to the ultraviolet. More excitingly, the hyperbolicity relation to anisotropic interband absorption in addition to the impressive dependency of the conduction band to the lattice constant along the out-of-plane direction provide the hyperbolicity tunability in these hexagonal structures under strain, doping, and alloying. Our findings not only suggest a large family of real hexagonal compounds as a unique class of materials for realization of the highly tunable broad band hyperbolicity but also offers an approach to search for new hyperbolic materials.
\end{abstract}

\maketitle


\section{Introduction}\label{sec:intro}

Thanks to relative permittivity manipulation through materials engineering, metamaterials have been the subject of intense interest in various science and engineering research communities aiming to conduct their interaction with electromagnetic and optical signals \cite{poddubny2013hyperbolic}. The relative permittivity, $\epsilon_{\rm 1}(\omega)$+i$\epsilon_{\rm 2}(\omega)$ is the main optical quantity that provides the description of the interaction between matter and electromagnetic waves. In hyperbolic metamaterials, $\epsilon_{\rm 1}(\omega)$ exhibits different signs along different directions revealing the extreme anisotropy of the electronic structure. Anisotropic optical materials exhibit numerous novel properties and potential applications, like negative refraction \cite{1, 2}, subwavelength imaging \cite{3, 4, 5} in the far-field, optical nanoscale cavity~\cite{6, 7}.

The electronic structure anisotropy is typically achieved in artificially engineered metamaterials suffering from complicated nanofabrication requirements, the high degree of interface electron scatterings, and the feature size of the components. Circumventing these limitations, natural hyperbolic materials provide the extended hyperbolic frequency window with low losses, high light confinement, and a larger photonics density of states~\cite{8, 9, 10, 11, 12, boos}. Heretofore several materials, like graphite, MgB$_{\rm 2}$, cuprate, ruthenate,and tetradymite have been predicted as natural hyperbolic materials~\cite{13, 14, 15, 16, 17},  while experimentally they were observed only in few of them, e.g., in MoO$_3$ surfaces\,\cite{bib:ma18,bib:zheng19}, structured hexagonal boron nitride\,\cite{bib:Li18} and T$_d\,$-WTe$_2$ thin films\,\cite{wang2020van}. Previous materials investigations reveal that the layered crystal structures exhibit excellent intrinsic out-of-plane optical anisotropy providing the required conditions for achieving natural hyperbolic materials~\cite{18, 19, 20}.

In the same way, a recent study shows that the layered nature of group-I nitrides with hexagonal P6/mmm ($\alpha$-type) crystal structure gives rise to an exceedingly strong anisotropy in the electronic structure~\cite{ebra}. The $\alpha$-Li$_{\rm 3}$N-type structure possesses a layered structure composed of alternating planes of the hexagonal Li$_{\rm 2}$N and pure Li$^{+}$-ions with weak interactions along the stacking direction. It should be noted that the similar behavior has been also reported in its counterparts; the high-temperature superconductor MgB$_2$ and AlB$_2$~\cite{mgb2, alb2}. In particular, the weak Coulombic interactions between adjacent layers compared with interactions in a layer leads to a less dispersive valence band maximum (VBM) along the $\Gamma$-A. It enhances the anisotropy of the electronic structures causing the highly anisotropic optical responses along the in- and out-of-plane directions of the crystals. Furthermore, the binary alkali pnictide A$_3$B (A = Li, Na; B = N, P, As, Sb, Bi) with hexagonal P6$_3$/mmc ($\beta$-type) structure meets the similar layered structure and nearly flat bands along the $\Gamma$-A which are responsible for the anisotropy of the electronic structure~\cite{p63mmc1, p63mmc2}. At this stage, we show, owing to these anisotropic responses, materials with $\alpha$ and $\beta$-type crystal structures belong to the rare class of natural hyperbolic materials via thoroughly investigation of the electronic and optical properties of $\alpha$-Li$_3$N. We emphasize that all the $\alpha$ and $\beta$-type structures have qualitatively the same physics as $\alpha$-Li$_3$N does (see supplemental materials).

The Li$_{\rm 3}$N, as a prominent member of this class, has been extensively studied as a distinguished superionic conductor while its optical properties remain silent. Using first principle calculations and Bethe-Salpeter Equation approach, we show that Li$_{\rm 3}$N is a natural hyperbolic in a broad spectral range from the visible to the ultraviolet spectrums. Interestingly, it exhibits type II hyperbolic response in lower and upper parts of the hyperbolic window while the middle one belongs to type I hyperbolic response. We find that the hyperbolicity in Li$_{\rm 3}$N is related to anisotropic interband absorption between the valence and conduction bands imposed by symmetry constraints and transition selection rules. 

As a standard feature of this new class of natural hyperbolic materials~\citep{20, alb2, p63mmc1}, the valence and conduction bandgap of Li$_{\rm 3}$N is highly impressionable and susceptible to the lattice constant {c}. These unique conditions provide the possibility of the hyperbolicity tunability in Li$_{\rm 3}$N with strain, doping and alloying. In fact, our calculations show that the hyperbolic window of Li$_{\rm 3}$N and the relative intensity of the sub-diffraction and diffraction-limited waves in transmission through Li$_{\rm 3}$N can be efficiently tuned by applying strain along the $z-$axis and doping in addition to using its ternary compounds like Li$_{\rm 2}$K(Na)N. Although Li$_{\rm 3}$N and Li$_{\rm 2}$K(Na)N possess the same structure, the larger ionic radius of K(Na) leads to a band crossing along the $z-$direction~\citep{ebra}, which extends the hyperbolic window in the visible regime highly demanding in light device applications. Beside Li$_{\rm 3}$N and the related alloys, the anisotropic electronic structure of $\alpha$ and $\beta$-type compounds gives rise to hyperbolic frequency windows for both type I and II as well offering hexagonal P6/mmm and P6$_3$/mmc layered structures as a unique class of materials for realization of the highly tunable broadband hyperbolicity and relevant device applications.

The structure of the paper is as follows. In Sec. II, we briefly introduce the simulation methods. Sec. III is devoted to the numerical results of the study, focusing on the electronic and optical properties of the $\alpha$-Li$_{\rm 3}$N crystal structure. Finally, we summarize our results in Sec. IV.

\section{Theoretical and Computational Methods}\label{sec:theory}
\begin{figure}[t]
	\centering
	\includegraphics[width=1.0\linewidth]{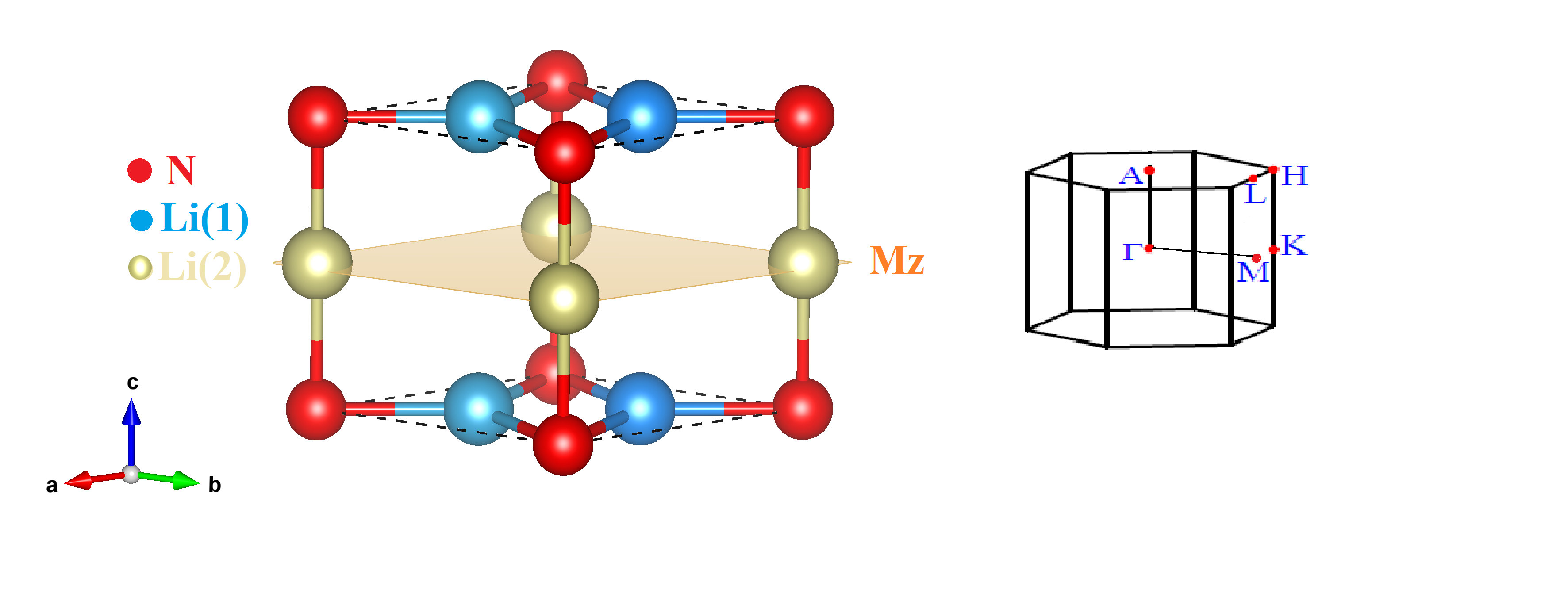}\\
	\caption{(Color online) The crystal structure (left panel) and the first Brillouin zone (right panel) of $\alpha$-Li$_{\rm 3}$N with P6/mmm symmetry. The mirror reflection plane is shown by M$_{\rm z}$. 
		\label{fig:1}}
\end{figure}

The electronic and linear optical properties of the insulator crystal structures (e.g. Li$_3$N) are calculated in the framework of density functional theory (DFT) and many-body perturbation theory. DFT is based on the accurate all-electron full-potential linearized augmented plane wave (FP-LAPW) method as implemented in Exciting code~\cite{22}. Ground-state properties are calculated using the hybrid density functional (HSE06)~\cite{23,24} in addition to PBE-type~\cite{25} generalized gradient approximation functional to meet adequately high accuracy requirements. The hole doping induced in the $\alpha$-Li$_{\rm 3}$N is simulated by shifting downward the Fermi level according to a standard rigid-band model. A 20$\times$20$\times$20 Monkhorst-pack k-point mesh is properly used in the computations. For Li$_{\rm 3}$N, the optical spectra are obtained by solving the Bethe-Salpeter equation (BSE); the equation of motion of the two-particle Green's function ~\cite{26, 27}. The matrix eigenvalue form of the BSE is thus given by~\cite{28, 29, 30, 31}
\begin{equation}
\sum_{\rm \nu^{\prime} c^{\prime}k^{\prime}} {\cal H}^{{\text {BSE}}}_{\bf \nu ck,\nu^{\prime} c^{\prime}k^{\prime}}{A}^{\lambda}_{\nu^{\prime} c^{\prime}{\bf k}^{\prime}} = {E}^{\lambda}{A}^{\lambda}_{\bf \nu ck}
\end{equation}
where $\nu$, $c$ and $k$ indicate the valence band, conduction band and k-points in the reciprocal space, respectively. Eigenvalues ${E}^{\lambda}$ and eigenvectors ${A}^{\lambda}_{\bf \nu ck}$ represent the excitation energy of the $j$th-correlated electron-hole pair and the coupling coefficients used to construct the exciton wave function, respectively. In the Tamm-Dancoff approximation the effective two-particle Hamiltonian for a spin-degenerate system is given by
\begin{equation} 
{\cal H}^{\text {BSE}} = {\cal H}^{\text {diag}} + 2{\cal H}^{x}+{\cal H}^{d}
\end{equation}
where the kinetic term ${\cal H}^{\text {diag}}$ is determined from vertical transitions between noninteracting quasiparticle energy and corresponds to the independent particle approximation. The screened Coulomb interaction, ${\cal H}^{d}$ and the bare electron-hole exchange, ${\cal H}^{x}$ are responsible for the formation of bound excitons.
Having calculated the eigenvalues and eigenvectors of the BSE, the long-wavelength limit of the imaginary part of the dielectric function $\epsilon_{\rm ii}(\omega)$ is given by ~\cite{28, 31}

\begin{equation} 
Im\epsilon_{\rm ii}(\omega)= \frac{8 {\pi}^2}{\Omega} \sum_{\rm \lambda} {{|t_{\rm \lambda}|}^2 \delta (\omega - E^{\lambda})}
\end{equation}
\begin{equation}  t_{\rm \lambda}=\sum_{\rm \nu c{\bf k}} {{A}^{\lambda}_{\nu c{\bf k}}\frac{<\nu k|\hat p|ck>}{\varepsilon^{\rm c}_{\bf k}-\varepsilon^{\nu}_{\bf k}} }
\end{equation}
where $\varepsilon^{\rm l}_{\bf k}$ with $l=c, \nu$ is the eigenvalue of the Kohn-Sham (KS) equation. 
The imaginary part of the DFT dielectric function shows the excitations in the system. The optical properties of semimetal structures like doped Li$_{\rm 3}$N, Li$_{\rm 2}$KN and Na$_{\rm 3}$N are considered using Time-dependent density-functional theory (TDDFT)~\cite{32}. The inverse dielectric matrix $\epsilon$ is related to the susceptibility $\chi$ (Ref. ~\cite{28, 33}) by the relation $\epsilon^{-1}(q,\omega)=1+v(q)\chi(q,\omega)$, where $v(q)$ is the bare Coulomb potential. In TDDFT the susceptibility $\chi(q,\omega)$ is obtained by making use of the linear response to TDDFT through the solution of the Dyson equation;
\begin{equation}
\chi(q,\omega)= {\chi}_{\rm 0}(q,\omega) + {\chi}_{\rm 0}(q,\omega)(v(q)+f_{xc}(q))\chi(q,\omega)
\end{equation}
where $\chi_{\rm 0}(q,\omega)$ is the KS susceptibility expressed in terms of the KS eigenvalues and eigenfunctions, while $f_{xc}(q)$ is the exchange-correlation kernel, for which we make use of the adiabatic local-density approximation (ALDA). The neglect of the $f_{xc}(q)$ term corresponds to the random-phase approximation (RPA). The macroscopic dielectric function $\epsilon _{M}$ is therefore defined as
\begin{equation} \epsilon _{M}({\bf q}', \omega)= \lim_{q \to 0} {\frac{1}{\epsilon^{-1}_{\bf G=0, G'=0}({\bf q}, \omega)}}
\end{equation}
where ${\bf q}' = {\bf q} + {\bf G}$ and ${\bf q}$ represents the momenta inside the first Brillouin zone.

\begin{figure*}[t]
\centering
\includegraphics[width=0.88 \textwidth]{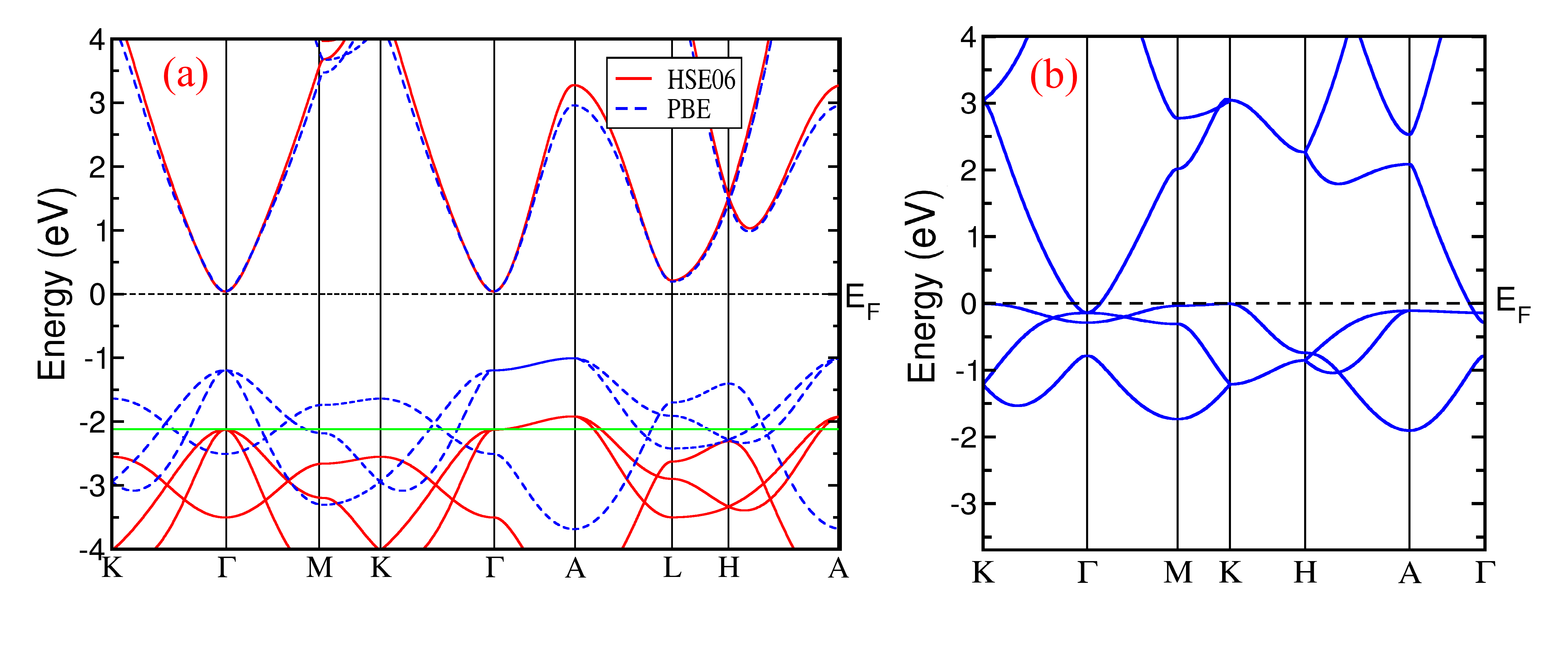} \\
\caption{(Color online) The calculated bulk electronic band structure of (a) Li$_{\rm 3}$N and (b) Na$_{\rm 3}$N (in PBE). The value of the Fermi energy for 0.1 holes/u.c is shown by a green solid line. The PBE gap of Li$_{\rm 3}$N is 1.04 eV while the gap is 2.12 eV within the HSE06. Except the bandgap value, the structures calculated by HSE06 are remarkably similar to those obtained within the PBE in the vicinity of the Fermi energy.
   \label{fig:ED}}
\end{figure*}

\section{\label{sec:level3} Numerical Results and Discussions}

Lithium nitrite ($\alpha$-Li$_{\rm 3}$N) crystallizes in the hexagonal P6/mmm structure ($\alpha$-Li$_{\rm 3}$N-type crystal structure) as shown in Fig.~\ref{fig:1}. The $\alpha$-Li$_{\rm 3}$N structure possesses a layered structure composed of alternating planes of the hexagonal Li$_{\rm 2}$N and pure Li$^{+}$-ions. In the Li$_{\rm 2}$N layers each N (0, 0, 0) is at the center of a symmetrical hexagon formed by the six neighboring Li-ions (1/3, 2/3, 0) and (2/3, 1/3, 0) in units of lattice vectors. From now on, these Li atoms denote as Li(1) and the Li atom above the N atoms consider as Li(2). In this structure, the three lithium atoms donating their 2s electrons to the nitrogen, resulting in Li$^{+}$-ions and an N$^{3-}$-ion. For the Li$_{\rm 3}$N crystal structure, we obtain the lattice constant {a}= {b}= 3.648 and {c}= 3.885 {\AA} which are completely consistent with the ones reported in experiment~\cite{34, 35, 36}.
\begin{figure*}[!htp]
	\includegraphics[width=0.95\linewidth]{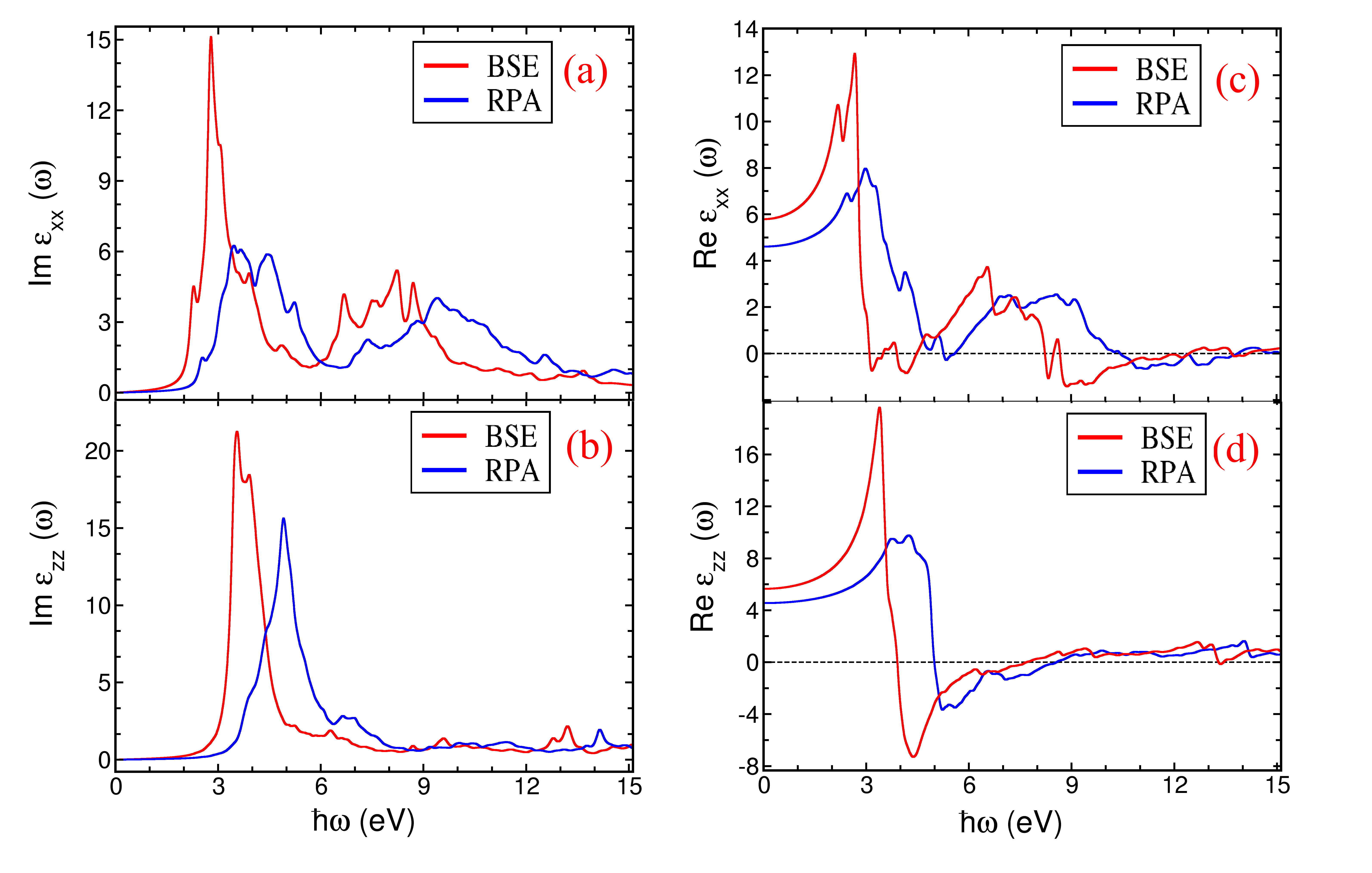}\\
	\caption{(Color online) The imaginary (a, b) and real (c, d) parts of the in-plane (top) and the out-of-plane (bottom) components of the macroscopic dielectric function Li$_{\rm 3}$N. The optical spectra of both the Im$\epsilon _{ii}(\omega)$ and Re$\epsilon _{ii}(\omega)$ show considerable anisotropy between the in-plane and out-of-plane components with distinct resonances in various directions. The polarization dependence originates from the structural anisotropy of Li$_{\rm 3}$N. The BSE results are qualitatively different from those obtained within the RPA especially in the lower part of the absorption. Moreover, the static dielectric constants along the in-plane and out-of-plane are 5.78 and 5.64 for Li$_{\rm 3}$N, respectively.
\label{fig:hf}}
\end{figure*}

The calculated orbital resolved band structure of Li$_{\rm 3}$N shows that the conduction band minimum (CBM) around the $\Gamma$ point is parabolic and mainly contributed by Li-s orbital while the valence band maximum (VBM) along the $\Gamma$-A is a doubly-degenerate band and mostly composed of N-p orbital.
The weak interactions along the $z$ direction lead to a less dispersive p$_{\rm z}$ VBM along the $\Gamma$-A and enhances the anisotropy of the electronic structures. This is the common feature of the $\alpha$ and $\beta$-type crystals and response structure for the anisotropic optical properties. 

As this novel natural hyperbolic class of materials contains also the semimetals with a Dirac node along the $\Gamma$-A, we depicte the band structure of Na$_{\rm 3}$N as a semimetal member example in Fig.~\ref{fig:ED}; providing a general overview of these electronic structures.
In spite of the crystal structure similarity between Na$_{\rm 3}$N and Li$_{\rm 3}$N, as shown in Fig.~\ref{fig:ED}, the CBM crosses the doubly-degenerate VBM along the $\Gamma$-A generating a triply degenerate nodal point (TDNP). However, due to similarities of optical properties of Li$_{\rm 3}$N and Na$_{\rm 3}$N, we focus on the Li$_{\rm 3}$N structure. The results for Na$_{\rm 3}$N and some other members are given in supplemental materials. In Li$_{\rm 3}$N, the PBE gap is 1.04 eV while the HSE06 gap is 2.12 eV otherwise the band structures calculated by HSE06 are remarkably similar to those obtained within the PBE in the vicinity of the Fermi energy (see Fig.~\ref{fig:ED}(a)). The dielectric tensor of the hexagonal Li$_{\rm 3}$N is diagonal and retains two independent components; $\epsilon _{xx}$ = $\epsilon_{yy}$ parallel to Li$_{\rm 2}$N plane (in-plane) and $\epsilon _{zz}$ perpendicular to the Li$_{\rm 2}$N plane (out-of-plane).

The BSE (RPA) computed optical spectra of Li$_{\rm 3}$N is illustrated in Fig.~\ref{fig:hf}. These calculations are performed by including HSE06 correction to PBE one. In general comparison, the optical spectra of both the Im$\epsilon _{ii}$ and Re$\epsilon _{ii}$ show considerable anisotropy between the in-plane and out-of-plane components with distinct resonances in various directions.
The polarization dependence of the optical absorption originates from the structural anisotropy of Li$_{\rm 3}$N.
In fact, the symmetry and space orientation of the valence and conduction bands would impose constraints on the optical transition selection rules ~\cite{37} causing the polarization dependence of the optical absorption.
The strength of the interband transition contribution to Im$\epsilon _{ii}$ is proportional to the product of the joint density of states (JDOS) and the electric dipole transition moment matrix $<\psi_{ \nu}|\hat{r}|\psi_{\rm c}>$, where $\psi_{\nu}$ and $\psi_{\rm c}$ are the wavefunctions of the valence and conduction bands, respectively and $r$ is the position operator.
\begin{figure*}
\centering
\includegraphics[width=0.88 \textwidth]{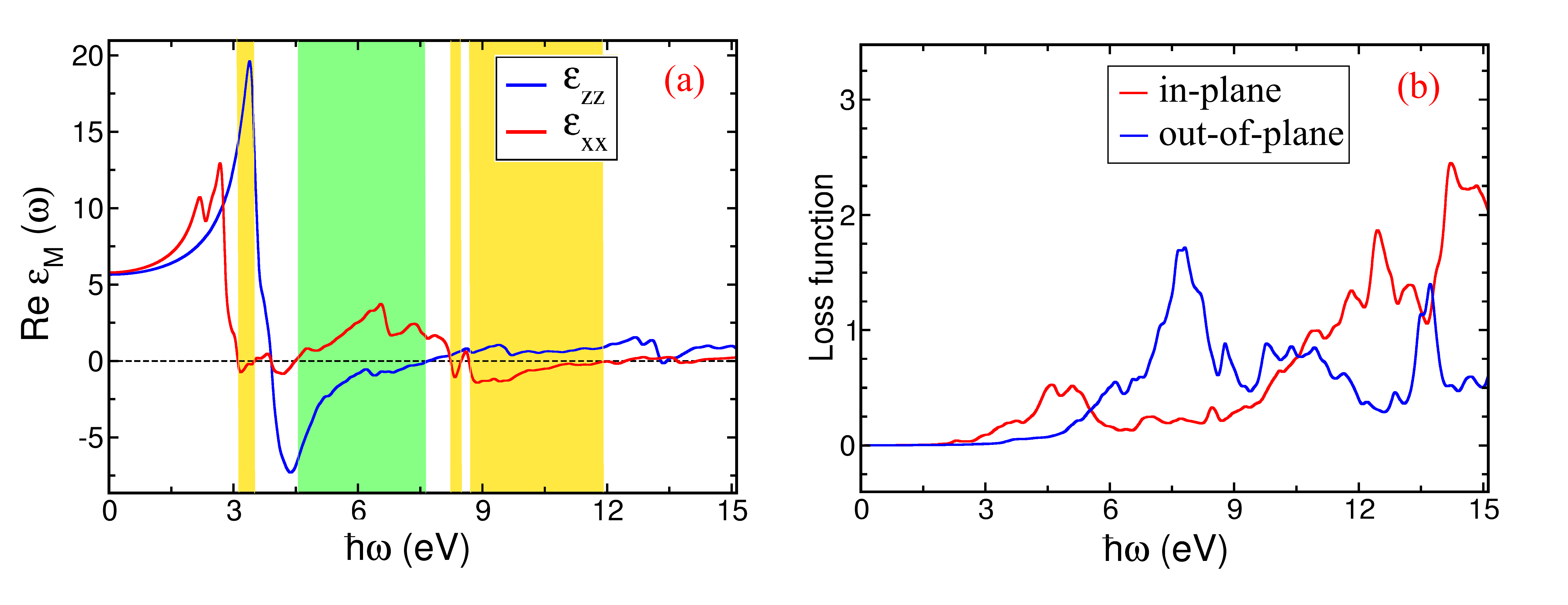} \\
\caption{(Color online) (a) The real part of the dielectric tensor and (b) the energy loss function to in-plane ($x-$polarized) and out-of-plane ($z-$polarized) incident light for Li$_{\rm 3}$N. The spectral region for type I and II hyperbolicity are marked with green and yellow background, respectively.  Three type-II hyperbolic frequency regimes in the visible region (from 3.1 eV to 3.54 eV) and ultraviolet (UV) regime including a narrow (from 8.23 eV to 8.5 eV) and a wide (from 8.67 eV to 12.48 eV) window.  Interband transition peaks in panel (b) are related to the peaks of Im$\epsilon _{ii}(\omega)$ while the plasmon peaks correspond to zeros of the Re$\epsilon _{ii}$. Notice that the plasmon excitation is located out of hyperbolic windows. \label{fig:E}}
\end{figure*}

As the VBM is mainly composed of N-p orbital and the CBM is constructed from Li-s orbital, the electric dipole transition moment matrix is $<s|\hat{r}|p>$. 
The nonzero elements, and the electric dipole transition are determined by the optical selection rules. At the $\Gamma$ point, the parity selection rule is fulfilled owing to the opposite parities of the VBM and CBM. As the crystal structure has a mirror symmetry, R$_{\rm z}$= $\sigma_{\rm z}$, the VBM and CBM have the same eigenvalues with respect to the mirror reflection operation R$_{\rm z}$ at the $\Gamma$ point.
\begin{figure*}[t]
\centering
\includegraphics[width=0.88 \textwidth]{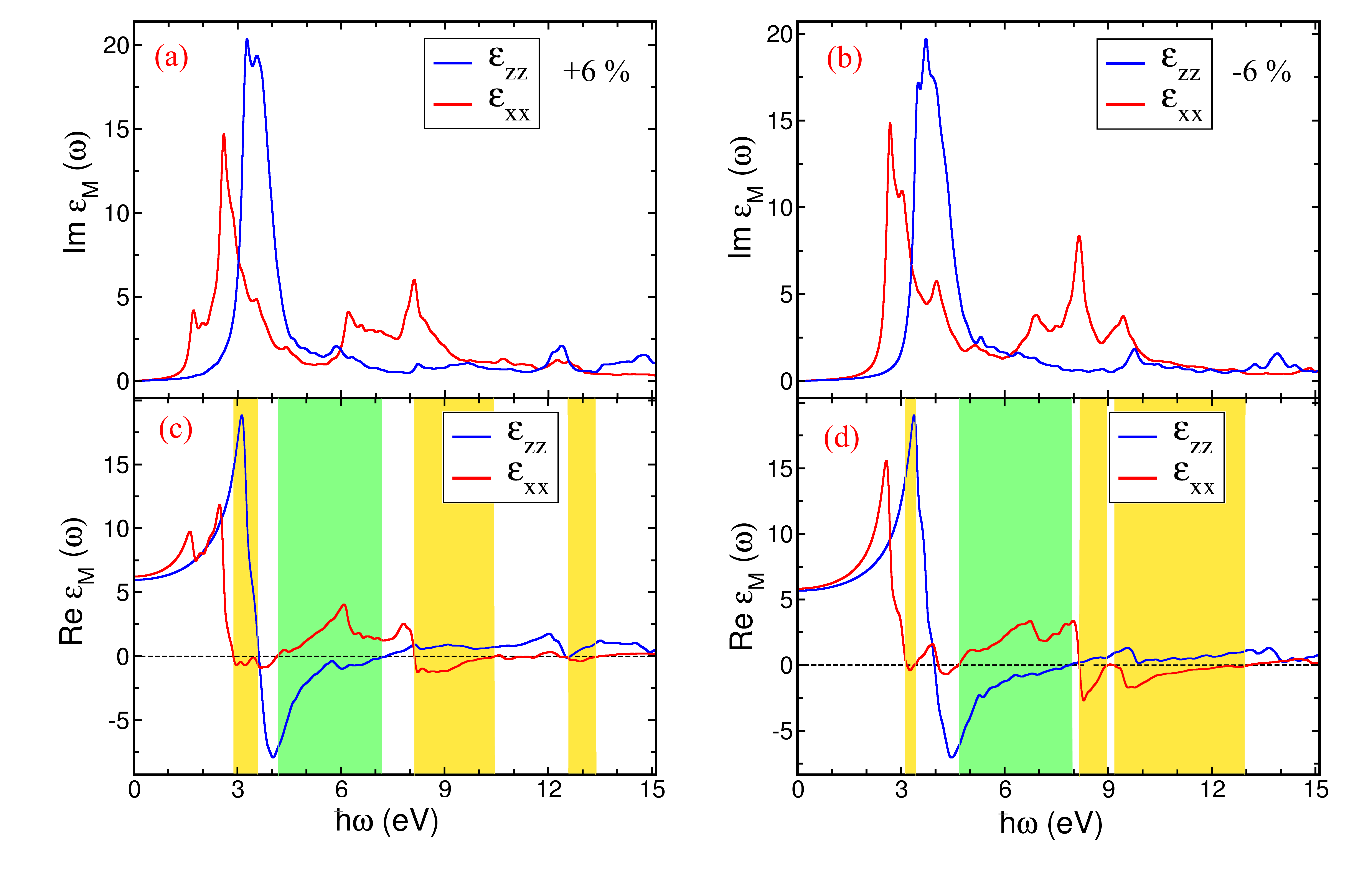} \\
\caption{(Color online) The imaginary (a, b) and real parts (c, d) of the dielectric tensor to the in-plane ($x$-polarized) and out-of-plane ($z$-polarized) incident light for +6$\%$ (a, c) and -6$\%$(b, d) strained Li$_{\rm 3}$N under tensile-strain along the $z$ axis. The spectral region for type I and II hyperbolicity are marked with the green and yellow background, respectively. The bandgap becomes smaller in +6$\%$ strained Li$_{\rm 3}$N (a, c) so the onset of the hyperbolic region moves to lower energy and gets inside the visible region. In addition, the hyperbolic windows are red-shifted and become broader (narrower) in the visible (UV) region by increasing (decreasing) the lattice constant $c$.
\label{fig:D}}
\end{figure*}
Therefore, at the $\Gamma$ point the electric dipole transition is symmetry allowed for the in-plane polarized light, while the transition in the out-of-plane direction is forbidden under the mirror symmetry R$_{\rm z}$. These symmetry constraints are responsible for an anisotropy between the in-plane and out-of-plane components of the dielectric tensors. The Im$\epsilon _{xx}$ has two main structures above 2 and 6 eV while the Im$\epsilon _{zz}$ has one structure above 2 eV and decays more rapidly. As a result, the Re$\epsilon _{xx}$ oscillates between positive and negative values providing the required conditions for achieving a hyperbolic regime. Interestingly, Li$_{\rm 3}$N is transparent in most parts of the UV spectrum for the out-of-plane polarization direction.
\begin{figure*}[t]
\centering
\includegraphics[width=0.88 \textwidth]{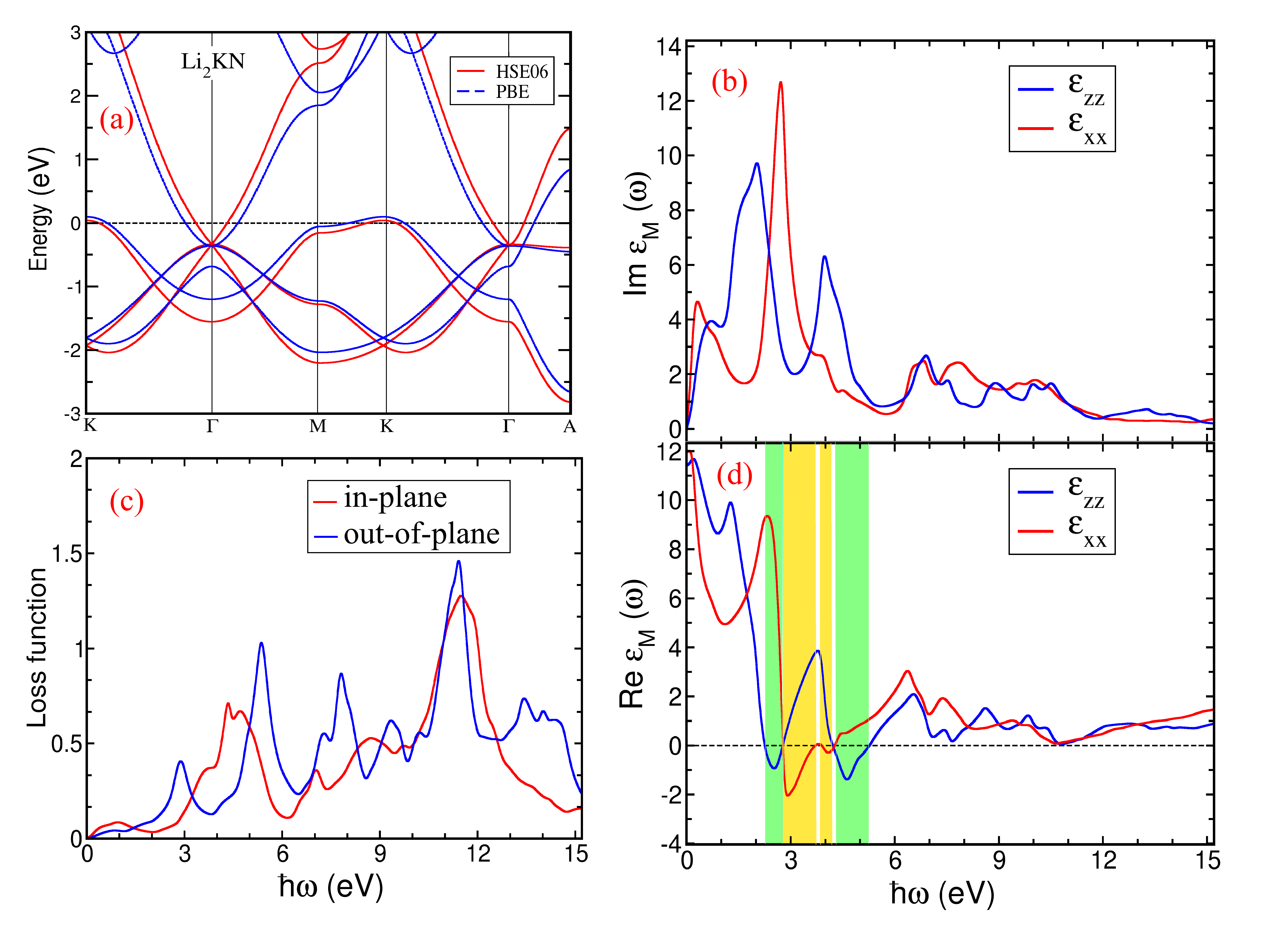} \\
\caption{(Color online) (a) The calculated bulk electronic band structure and (c) the energy loss function to the in-plane ($x$-polarized) and out-of-plane ($z$-polarized) incident light for Li$_{\rm 2}$KN. (b) The imaginary and (d) real parts of the dielectric tensor to the in-plane ($x$-polarized) and out-of-plane ($z$-polarized) incident light for Li$_{\rm 2}$KN. The spectral region for type I and II hyperbolicity are marked with the green and yellow background, respectively. Type-I hyperbolic frequency regime is broader than type-II hyperbolic window with a hyperbolicity onset in the visible region at 2.3 eV. The transition from type I to type II occurs at 2.79 eV where Re$\epsilon_{\rm xx}(\omega)$ and Re$\epsilon_{\rm zz}(\omega)$ vanish with an opposite sign slope. Furthermore, the loss function components show small values meaning that there is no energy loss when the material is hyperbolic.
\label{fig:F}}
\end{figure*}
As shown in Fig.~\ref{fig:hf}, the BSE results are qualitatively different from those obtained within the RPA especially in the lower part of the absorption where the absorption onset is dominated by two intense peaks formed by bound excitons with large oscillator strengths. The analysis of the k-space distribution of the exciton weights shows that the first excitation originates from the VBM to the conduction band at the $\Gamma$ point where the lowest-energy excitation is produced by the transition between the highest occupied band and the lowest-unoccupied band at the $\Gamma$ point. The next excitations are mainly constituted by transitions between the first (second) highest occupied band and the lowest unoccupied band along the path $\Gamma$-A ($\Gamma$-K), close to the $\Gamma$ point.
Moreover, the static dielectric constants along the in-plane and out-of-plane are 5.78 and 5.64 for Li$_{\rm 3}$N, respectively.
In considering the Re$\epsilon _{ii}$ of Li$_{\rm 3}$N, Fig.~\ref{fig:hf}(c, d), shows the negative values of the out-of-plane component of the real part of the dielectric function only from 3.94 up to 7.66 eV which correspond to a high reflectivity region. Most excitingly, the in-plane component of Re$\epsilon _{ii}$ is mostly positive in this region while it goes through a high reflectivity region in outside of the region. As the in-plane and out-of-plane components of Re$\epsilon _{ii}$ cross zero at different frequencies, the Li$_{\rm 3}$N fulfills the necessary conditions for type I (Re$\epsilon _{xx}$ $>$ 0 and Re$\epsilon _{zz} $ $<$ 0) and II (Re$\epsilon _{xx}$ $<$ 0 and Re$\epsilon _{zz}$ $>$ 0) indefinite permittivity media.

There are three clear type-II hyperbolic frequency regimes in the visible region (from 3.1 eV to 3.54 eV) and ultraviolet (UV) regime including a narrow (from 8.23 eV to 8.5 eV) and a wide (from 8.67 eV to 12.48 eV) window as shown in Fig.~\ref{fig:E}(a). The wide type-I hyperbolic frequency regime is located between 4.51 eV and 7.66 eV.
Like type-I hyperbolic window supports propagating diffraction-limited and sub-diffraction waves while the type II hyperbolic region only supports the latter one. The relative intensity of the sub-diffraction and diffraction-limited waves in transmission through Li$_{\rm 3}$N can be tuned by modifying the type of the hyperbolic window providing a unique platform to the definitive experimental study of the exotic phenomena like the anti-cutoff~\cite{38}. An ideal hyperbolic material should be nearly lossless in the hyperbolic window.

We show, on the other hand, the energy loss function of Li$_{\rm 3}$N in Fig.~\ref{fig:E}(b).
The energy loss function, L($\epsilon(q,\omega)$) = Im (-1/$\epsilon(q,\omega)$), which determined the energy loss of a fast electron moving across a medium, is a complicated mixture of interband transitions and plasmons. Interband transition peaks are related to the peaks of Im$\epsilon _{ii}$ while the plasmon peaks correspond to zeros of the Re$\epsilon _{ii}$. The plasmon peak of the loss function is larger if Re$\epsilon _{ii}$ is zero and Im$\epsilon _{ii}$ is a significantly small value. As shown in Fig.~\ref{fig:E}(b), both the components of the loss function show negligible values in the entire hyperbolic energy window except a broad plasmon peak with a small intensity above 7.5 eV. This plasmon excitation is actually located out of hyperbolic windows. Therefore, Li$_{\rm 3}$N is a low-loss in hyperbolic frequency regimes making it an ideal natural hyperbolic material.
The origins of the hyperbolic response regimes can be detected by considering the optical transitions between the valence and conduction bands in the hyperbolicity windows, shown in the electronic band structure in Fig.~\ref{fig:hf}.

\begin{figure*}[t]
\centering
\includegraphics[width=0.88 \textwidth]{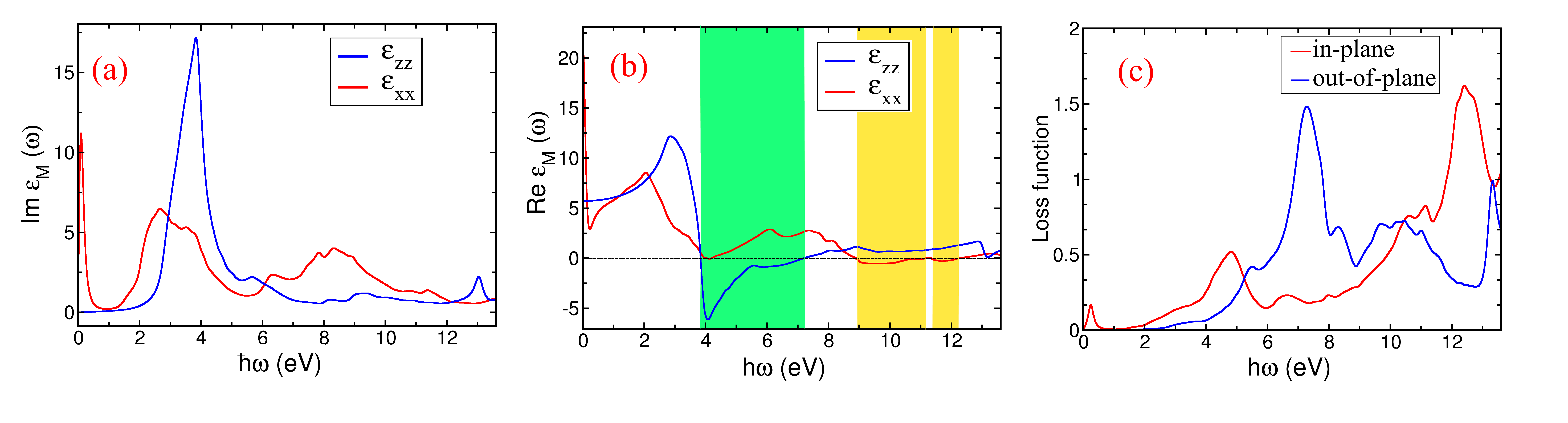} \\
\caption{(Color online) (a) The imaginary and (b) real parts of the dielectric tensor to the in-plane ($x$-polarized) and out-of-plane ($z$-polarized) incident light for hole-doped (0.1 holes/u.c.) Li$_{\rm 3}$N. The spectral region for type I and II hyperbolicity are marked with the green and yellow background, respectively. (c) The energy loss function for in-plane and out-of-plane incident light. Under hole doping the $x-$polarized interband absorption above 2 and 6 eV becomes weaker and broader than corresponding ones in pristine Li$_{\rm 3}$N. The negative values of the in-plane component of the real part of the dielectric function is less pronounced in hole-doped. The loss function reflects the anisotropic electronic structures. Up to 5 eV the out-of-plane component of the loss function is nearly zero while the in-plane component contains two lesser peaks. 
\label{fig:H}}
\end{figure*}

Owing to the Kramers-Kronig relation~\cite{kramer} between the real and imaginary parts of the dielectric function, the high oscillator strength resonances of Im$\epsilon _{xx}$ and Im$\epsilon _{zz}$ at ~ 2.8 eV and 3.7 eV are responsible for the negative permittivity in Re$\epsilon _{xx}$ and Re$\epsilon _{zz}$ at ~ 3.1 eV and ~ 3.9 eV in Fig.~\ref{fig:E}(a). These intense peaks in Im$\epsilon _{xx}$ and Im$\epsilon _{zz}$ are related to the interband transitions between the valence and conduction bands along the $\Gamma$-A and the L-A (see Fig.~\ref{fig:ED}). The band structure of the system determines the occupied and unoccupied bands nearly parallel to each other in these domains and therefore have the same dispersions, resulting in a constant energy difference between the conduction and valence bands (${|\bigtriangledown_{{\bf k}}(E_{\rm c}-E_{\nu})|}=0$)~\cite{39}.
As the JDOS can be expressed in terms of ${|\bigtriangledown_{{\bf k}}(E_{\rm c}-E_{\nu})|}$ ~\cite{39}:
\begin{equation}
\frac{1}{{2 \pi}^2}\int \frac{d{S_{k}}}{|\bigtriangledown_{{\bf k}}(E_{\rm c}-E_{\nu})|}
\end{equation}
where ${S_{k}}$ is the constant energy surface defined by ${E_{\rm c}-E_{\nu}}={\text {const.}}$), thereby, the JDOS and absorption are significantly large along the $\Gamma$-A and L-A directions.

The same as before, the polarization dependence of the negative permittivity window (absorption) along the $\Gamma$-A and L-A directions is dictated by symmetry constraints and transition selection rules. The point group along the high symmetry lines, the $\Gamma$-A and L-A, are D$_{\rm 6h}$ and C$_{\rm 2v}$ which have horizontal mirror plane $\sigma_{\rm d}$ (R$_{\rm z}$ = $\sigma_{\rm z}$) and vertical mirror plane $\sigma_{\nu}$, respectively. The VBM and CBM have the same eigenvalues with respect to the mirror reflection operation R$_{\rm z}$ along the $\Gamma$-A direction while they maintain the same parity with respect to the vertical mirror plane $\sigma_{\rm v}$ along the L-A. These constraints make the $z-$polarized transitions ($x-$polarized transitions) forbidden along the $\Gamma$-A (L-A) causing the strong anisotropic dielectric tensors which represent the required condition for the hyperbolicity.

Furthermore, under tensile-strain along the $z-$axis, the bandgap of Li$_{\rm 3}$N becomes smaller so the onset of the hyperbolic region moves to lower energy and gets inside the visible region as described in Fig.~\ref{fig:D}. In addition, the hyperbolic windows are red-shifted and become broader (narrower) in the visible (UV) region by increasing (decreasing) the lattice constant $c$. Inversely, under the compressive strain, the spectral range of hyperbolicity moves to the UV region and becomes smaller due to the bandgap increasing.

Having extensively explored the optical properties of the Li$_{\rm 3}$N, we now investigate the possible existence of the hyperbolicity in related alloys like Li$_{\rm 2}$KN crystal structures. The lattice constants and internal coordinates of Li$_2$KN were fully optimized and the obtained results ({a}= {b}= 3.713 and {c}= 5.701 {\AA}) are completely consistent with the previous report~\cite{schon2000investigation}. Our calculations show that replacing Li-ion on top of the N with Potassium (K) atoms reduces the interactions between the N-p$_{\rm z}$ and K-s orbitals leading to the reversal of the band ordering between the conduction and valence bands at the center of the first BZ of Li$_{\rm 3}$N (see Fig.~\ref{fig:F}). In comparison to Li$_{\rm 3}$N, the spectral range of the hyperbolicity are smaller and located below 6 eV in the Li$_{\rm 2}$KN. In Li$_{\rm 2}$KN, type-I hyperbolic frequency regime is broader than type-II hyperbolic window with a hyperbolicity onset in the visible region at 2.3 eV which is highly demanding in light device applications.
The transition from type I to type II indefinite permittivity media occurs at 2.79 eV where Re$\epsilon_{\rm xx}$ and Re$\epsilon_{\rm zz}$ become zero with an opposite sign slope. The same scenario happens at 4.2 eV and type II transforms into type I indefinite permittivity media. The type I hyperbolic frequency regime extends up to 5.3 eV. This interesting arrangement of various types of hyperbolic regimes makes Li$_{\rm 2}$KN an ideal tunable natural indefinite permittivity media where the type of hyperbolicity can be selected by tuning the energy of light. Furthermore, the loss function components show small values which means that there is almost no energy loss when the material is hyperbolic.

Finally, we consider the tunability of the optical properties of Li$_{\rm 3}$N by regulating its Fermi level with doping. We consider 0.1 holes per unit cell by setting the Fermi energy to 0.17 eV below the VBM (see Fig.~\ref{fig:ED}). As the Fermi level crosses two valence bands along the perpendicular to the $z-$direction, an intense peak appears in Im$\epsilon _{xx}$ at $\sim$ 0.1 eV (Fig.~\ref{fig:H}). In comparison to pristine Li$_{\rm 3}$N, under hole doping the $x-$polarized interband absorption above 2 and 6 eV becomes weaker and broader than corresponding ones in pristine Li$_{\rm 3}$N. As a result, the negative values of the in-plane component of the real part of the dielectric function is less pronounced in hole-doped Li$_{\rm 3}$N causing the blue shift in type-II hyperbolic regimes. Our results show that the type-II hyperbolic is just appeared in the UV region (from 8.9 to 12.1). However, the main intense peak of Im$\epsilon _{zz}$ is preserved and red-shifted in hole-doped Li$_{\rm 3}$N. Therefore, the type-I hyperbolic frequency regime is located between 3.8 eV and 7.2 eV. The loss function of a hole-doped Li$_{\rm 3}$N reflects perfectly the anisotropic electronic structures. Up to 5 eV the out-of-plane component of the loss function is nearly zero while the in-plane component has two small peaks. From 6 to 10 eV, the situation gets reversed and the out-of-plane component of the loss function exhibits a small peak in the vanishing region of the in-plane component. However, neglecting the peak around 7 eV, the loss function is negligible up to 11 eV.

\section{DISCUSSION AND CONCLUSION}
Natural two-dimensional materials with anisotropic optical response are considered promising candidates for hosting hyperbolic isofrequency surface. In addition, two-dimensional materials are characterized with highly confined and low-loss polaritons, that can be additionally tuned with field effect techniques as well as with strain. In the present work, we provide a detailed ab initio study on an anisotropic optical response in the hexagonal crystal structures. Using density functional theory and symmetry analysis, we have proposed $\alpha$- and $\beta$-type hexagonal layered crystal structure as a new class of hyperbolic materials where the hyperbolicity is arisen from the weak Coulombic interactions between adjacent layers compared with interactions within a specific layer leads in addition to symmetry constraints and transition selection rules. A comprehensive study of optical properties introduces this class of materials as a new natural hyperbolic material in which the type of hyperbolicity can be selected by changing the frequency of light. More excitingly, the hyperbolicity in these materials can be tuned using lattice strain along the $z$-axis and doping as well. Furthermore, the many-body effects on the dielectric functions are discussed. We have also shown this class of materials are a low-loss in hyperbolic frequency regimes making them ideal natural hyperbolic materials. Our findings here can be explored by current experiments.

Ultimately, we would like to emphasize potential viable applications of $\alpha$-Li$_{\rm 3}$N-( $\beta$-Li$_{\rm 3}$N-) type hexagonal layered crystal structures in comparison with layered transition metal dichalcogenides (TMDs)~\cite{gjerding2017}. First of all, as discussed in this paper, this new class of hyperbolic materials exhibits both type I and type-II hyperbolicity over the wide range of frequencies which rarely found in nature~\cite{gjerding2017}. As a result, the type of hyperbolicity can be selected by tuning the energy of light. Secondly, a previous study shows that the low-losses as main responsible of a large Purcell factor in the hyperbolicity regime only meet at the metallic TMDs where the conduction bands being sufficiently separated from other bands by finite energy bandgaps. However, our calculations show that $\alpha$-Li$_{\rm 3}$N-( $\beta$-Li$_{\rm 3}$N-) type crystal structures, both metallic and semiconductor ones, are nearly lossless in the hyperbolic windows. Most importantly, in $\alpha$-Li$_{\rm 3}$N-( $\beta$-Li$_{\rm 3}$N-) type crystal structures the impressive dependency of the conduction band to the lattice constant $c$ provide the hyperbolicity tunability in these hexagonal structures under strain, doping, and alloying which are more plausible than heterostructures of standard TMDs.
\section{Acknowledgments}

A. E is supported by Iran Science Elites Federation and R. A is supported by the Australian Research Council Centre of Excellence in Future Low-Energy Electronics Technologies (project number CE170100039).

\nocite{apsrev41Control}
\bibliographystyle{apsrev4-1}
\bibliography{H}

\end{document}